\documentclass[letterpaper, 10 pt, conference]{ieeeconf}

\IEEEoverridecommandlockouts   
\ifx\pdfoutput\undefined
\usepackage{graphicx}
\else
\usepackage[pdftex]{graphicx}
\fi

\usepackage{latexsym,amssymb,amsmath,ifthen,epic}

\newcommand{\ignore}[1]{}

\newcommand{\visual}{}

\newcommand{\Fig}[1]{Fig.~\ref{#1}}

\newcommand{\eqdef}{\stackrel{\scriptscriptstyle\bigtriangleup}{=} }
\newcommand{\eqm}[1]{\parbox{2em}{\centering $#1$}}  

\newcommand{\argmin}{\operatornamewithlimits{argmin}}

\newcommand{\muleft}{\protect\overset{\,\leftarrow}{\mu}}
\newcommand{\muright}{\protect\overset{\;\rightarrow}{\mu}}

\newcommand{\murighttrue}{\muright_{\!\text{true}}}
\newcommand{\murightapprox}{\muright_{\!\text{approx}}}

\newcommand{\muleftb}{\muleft_{\!\text{b}}}
\newcommand{\murightb}{\muright_{\!\text{b}}}
\newcommand{\muleftg}{\muleft_{\!\text{g}}}
\newcommand{\murightg}{\muright_{\!\text{g}}}
\newcommand{\murightgM}{\muright_{\!\text{gM}}}
\newcommand{\murightgs}{\muright_{\!\text{gs}}}

\newcommand{\mleft}{\protect\overset{\,\leftarrow}{\text{m}}}
\newcommand{\mright}{\protect\overset{\,\rightarrow}{\text{m}}}

\newcommand{\mleftg}{\mleft_{\text{g}}}
\newcommand{\sigmaleftg}{\overleftarrow{\sigma_{\text{g}}^2}}
\newcommand{\mleftb}{\mleft_{\text{b}}}

\newcommand{\mrightgs}{\mright_{\text{gs}}}
\newcommand{\sigmarightgs}{\overrightarrow{\sigma_{\text{gs}}^2}}
\newcommand{\mrightgM}{\mright_{\text{gM}}}
\newcommand{\sigmarightgM}{\overrightarrow{\sigma_{\text{gM}}^2}}
\newcommand{\mrightg}{\mright_{\text{g}}}
\newcommand{\sigmarightg}{\overrightarrow{\sigma_{\text{g}}^2}}
\newcommand{\mrightb}{\mright_{\text{b}}}
\newcommand{\sigmarightb}{\overrightarrow{\sigma_{\text{b}}^2}}

\newcommand{\mtrue}{\text{m}_{\text{true}}}
\newcommand{\sigmatrue}{\sigma_{\text{true}}^2}
\newcommand{\mg}{\text{m}_{\text{g}}}
\newcommand{\sigmag}{\sigma_{\text{g}}^2}

\newcommand{\cent}[1]{\makebox(0,0){#1}}
\newcommand{\pos}[2]{\makebox(0,0)[#1]{#2}}

\begin{document}



\title{\LARGE \bf
A General Computation Rule for Lossy Summaries/Messages\\ with Examples from Equalization}


\author{Junli Hu, Hans-Andrea Loeliger, Justin Dauwels, and Frank Kschischang%
\thanks{Junli Hu and Hans-Andrea Loeliger are with 
  the Dept.\ of Information Technology and Electrical Engineering, 
  ETH Zurich, CH-8092 Zurich, Switzerland.}%
\thanks{Justin Dauwels is with the 
  Amari Research Unit of the RIKEN Brain Science Institute, 
  2-1 Hirosawa, Wako-shi, Saitama-ken, 351-198, Japan.}%
\thanks{Frank Kschischang is with the
  Electrical and Computer Engineering Dept., University of Toronto, 
  Ontario M4S 3G5, Canada.}%
}

\maketitle

\begin{abstract}
Elaborating on prior work by Minka, 
we formulate a general computation rule for lossy messages. 
An important special case (with many applications in communications) 
is the conversion of ``soft-bit'' messages to Gaussian messages. 
By this method, the performance of a Kalman equalizer is improved, 
both for uncoded and coded transmission.
\end{abstract}

\section{Introduction}
We consider message passing algorithms in factor graphs \cite{KFL:fg2000}, \cite{Lg:ifg2004}. 
If the factor graph has no cycles, the messages computed by 
the basic sum-product and max-product algorithms are \emph{exact} summaries 
of the subgraph behind the corresponding edge. 
However, in many applications (especially with continuous variables), 
complexity considerations suggest, or even dictate, the use of 
\emph{approximate} or \emph{lossy} summaries. 
For example, it is customary to use Gaussian messages even in cases where the ``true'' 
(sum-product or max-product) messages are not Gaussian,  
or to use scalar (i.e., single-variable) messages 
instead of multi-dimensional (i.e., multi-variable) messages. 

In this paper, we first formulate a general message update rule for lossy summaries/messages 
that is a nontrivial generalization of the standard sum-product or max-product rules. 
This rule was in essence proposed by Minka \cite{Mi:thesis2001}, \cite{Mi:ep}, 
but our general formulation of it may not be obvious from Minka's work. 

We then focus on one particular application: the conversion of binary (``soft-bit'') messages 
into Gaussian messages, which has many uses in communications. 
For our numerical examples, we then further focus on 
equalization: we give simulation results for an iterative Kalman equalizer 
both for a linear FIR (finite impulse response) channel 
and for a linear IIR (infinite impulse response) channel. 
For uncoded transmission,
the new algorithm almost closes the gap between the BJCR algorithm 
and the LMMSE (linear minimum mean squared error) equalizer;
for coded transmission, the new algorithm 
improves the performance of the iterative Kalman equalizer
at very little additional cost. 

It should be noted that the new message computation rule yields iterative algorithms 
even for cycle-free graphs. We also note that some sort of damping is usually 
required to stabilize the algorithm. 

In this paper, we will use Forney-style factor graphs as in \cite{Lg:ifg2004} 
where edges represent variables and nodes represent factors.

\section{A General Computation Rule for Lossy Messages/Summaries}
\label{sec:GeneralMinkaRule}
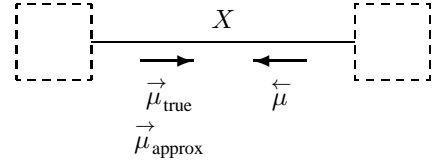
\begin{figure}
\begin{center}
\begin{picture}(55,15)(0,0)
%
\put(0,5){\dashbox(10,10){}}       
\put(10,10){\line(1,0){35}}        \put(27.5,13){\cent{$X$}}
{\thicklines
 \put(16.5,7.5){\vector(1,0){7}}   \put(20,3){\cent{$\murighttrue$}}
                                   \put(20,-3){\cent{$\murightapprox$}}
 \put(38.5,7.5){\vector(-1,0){7}}  \put(35,3){\cent{$\muleft$}}
}
\put(45,5){\dashbox(10,10){}}
\end{picture}
\vspace{5mm}
\caption{\label{fig:GenRuleSetup}%
Lossy message $\murightapprox$ along a general edge/variable $X$.}
\end{center}
\end{figure}

Consider the messages along a general edge (variable) $X$ 
in some factor graph as illustrated in \Fig{fig:GenRuleSetup}. 
Let $\murighttrue(x)$ be the ``true'' sum-product or max-product message 
which we want (or need) to replace by a message $\murightapprox(x)$ 
in some prescribed family of functions (e.g., Gaussians). 
In such cases, 
most writers (including these authors) used to compute $\murightapprox$ 
as some approximation of $\murighttrue$. 
However, the semantics of factor graphs suggests another approach. 
Note that the factor graph of \Fig{fig:GenRuleSetup} represents 
the function 
\begin{equation} \label{eqn:TrueGlobalFunction}
f(x) \eqdef \murighttrue(x) \muleft(x),
\end{equation}
which the replacement of $\murighttrue$ by $\murightapprox$ will change into 
\begin{equation} \label{eqn:ApproxGraph}
\tilde f(x) \eqdef \murightapprox(x) \muleft(x).
\end{equation}
It is thus natural to first compute 
\begin{equation} \label{eqn:GenRuleApprox}
\tilde f(x) = \text{some approximation of~} \murighttrue(x) \muleft(x)
\end{equation}
and then to compute $\murightapprox$ from (\ref{eqn:ApproxGraph}).
The approximation in (\ref{eqn:GenRuleApprox}) must be chosen 
so that solving (\ref{eqn:ApproxGraph}) for $\murightapprox$ 
yields a function in the prescribed family.

Important special cases of this general approach (including the Gaussian case) 
were proposed as ``expectation propagation'' in \cite{Mi:thesis2001} and \cite{Mi:ep}. 

The choice of a suitable approximation in (\ref{eqn:GenRuleApprox}) 
will, in general, depend on the application. For many applications, 
a natural approach (proposed and pursued by Minka) 
is to minimize the Kullback-Leibler divergence: 
\begin{equation} \label{eqn:MinDivergence}
\tilde f = \argmin_{f' \text{~in chosen family}} D(f \| f'). 
\end{equation}

In this paper, the approximate messages will always be Gaussian. 
However, other families of functions can be used. 
For example, multivariable messages 
with a prescribed Markov-chain structure were used in \cite{DLMO:Allerton2004c}; 
with hindsight, the update rule for such messages that was proposed in \cite{DLMO:Allerton2004c} 
is indeed an example of the general scheme described here.
A related idea was proposed in \cite{MiQi:tsep2003c}.

\section{Converting Soft-Bit Messages to Gaussian Messages}
\label{sec:BinToGaussian}
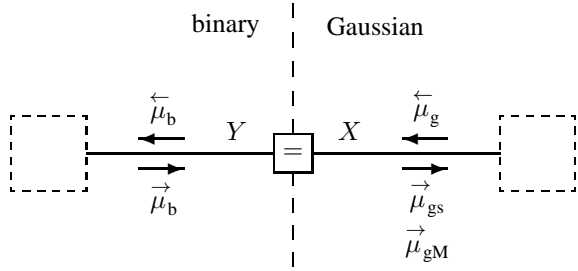
\begin{figure}
\begin{center}
\begin{picture}(75,35)(0,0)
%
\put(0,10){\dashbox(10,10){}}
\put(10,15){\line(1,0){25}}
{\thicklines
 \put(23,17){\vector(-1,0){6}}    \put(20,21){\cent{$\muleftb$}}
 \put(17,13){\vector(1,0){6}}     \put(20,9){\cent{$\murightb$}}
}
\put(30,18){\cent{$Y$}}
\dashline{2}(37.5,17.5)(37.5,35)    \put(33,32){\pos{r}{binary}}
                                    \put(42,32){\pos{l}{Gaussian}}
\put(35,12.5){\framebox(5,5){$=$}}
\dashline{2}(37.5,0)(37.5,12.5)
\put(40,15){\line(1,0){25}}
\put(45,18){\cent{$X$}}
{\thicklines
 \put(58,17){\vector(-1,0){6}}    \put(55,21){\cent{$\muleftg$}}
 \put(52,13){\vector(1,0){6}}     \put(55,9){\cent{$\murightgs$}}
                                  \put(55,3){\cent{$\murightgM$}}
}
\put(65,10){\dashbox(10,10){}}
\end{picture}
\end{center}
\vspace{-1mm}
\caption{\label{fig:BinToGaussian}%
Converting a soft-bit message $\murightb$ into a Gaussian message $\murightgs$ or $\murightgM$.}
\end{figure}

We will now apply the general scheme of the previous section 
to the conversion of messages defined on the finite alphabet $\{+1, -1\}$ 
into Gaussian messages. 
The setup is shown in \Fig{fig:BinToGaussian}, 
which is (a part of) a factor graph with an equality constraint 
between the real variable $X$ and the $\{+1,-1\}$-valued variable $Y$. 
(The equality constraint node in \Fig{fig:BinToGaussian} may formally be viewed 
as representing the factor $\delta(x-y)$, which is to be understood 
as a Dirac delta in $x$ and a Kronecker delta in~$y$.)  
The messages $\murightb$ and $\muleftb$ are defined on the finite 
alphabet $\{+1, -1\}$ and the messages $\muleftg$, $\murightgs$, and $\murightgM$ are 
Gaussians; $\murightgs$ denotes the standard Gaussian approximation and $\murightgM$ 
denotes the alternative Gaussian approximation due to Minka, as will be detailed below. 

Let us first recall the conversion of Gaussian messages into soft-bit messages. 
Let $\mleftg$ and $\sigmaleftg$ be the mean and the variance, respectively, 
of $\muleftg$. 
The (lossless) conversion from $\muleftg$ to $\muleftb$ is an immediate 
and standard application of the sum-product (or max-product) rule 
\cite{KFL:fg2000}, \cite{Lg:ifg2004}:
\begin{equation}  \label{eqn:muleftb}
\left( \begin{array}{c}
          \muleftb(+1) \\
          \muleftb(-1) 
       \end{array} \right)
\propto
\left( \begin{array}{c}
          \muleftg(+1) \\
          \muleftg(-1)
       \end{array} \right);
\end{equation}
in the standard logarithmic representation, this becomes 
\begin{equation}
\ln \frac{\muleftb(+1)}{\muleftb(-1)} 
= \frac{2\mleftg}{\sigmaleftg}.
\end{equation}

We now turn to the more interesting lossy conversion of 
$\murightb$ into a Gaussian. 
Let $\mrightb$ and $\sigmarightb$ be 
the mean and the variance, respectively, of $\murightb$, 
which are given by
\begin{align}
\mrightb  &\eqm{=}     \frac{\murightb(+1) - \murightb(-1)}{\murightb(+1) + \murightb(-1)} \label{eqn:MeanOfMrightb}\\
\sigmarightb &\eqm{=}  1 - (\mrightb)^2.
\end{align}

The traditional approach forms the Gaussian message $\murightgs$ 
(with mean $\mrightgs$ and variance $\sigmarightgs$) 
from the mean and the variance of $\murightb$:
\begin{equation} \label{eqn:StandardGaussianApprox}
\mrightgs = \mrightb \text{~~~and~~~} \sigmarightgs = \sigmarightb. 
\end{equation}
The approach of Section~\ref{sec:GeneralMinkaRule} yields 
another Gaussian message $\murightgM$ 
(with mean $\mrightgM$ and variance $\sigmarightgM$) 
as follows. 
In \Fig{fig:BinToGaussian}, 
the true global function corresponding to (\ref{eqn:TrueGlobalFunction}) is
\begin{equation}
\delta(x-1) \murightb(+1) \muleftg(+1) + \delta(x+1) \murightb(-1) \muleftg(-1)
\end{equation}
which (when properly normalized) has mean 
\begin{equation} \label{eqn:mtrue}
\mtrue = \frac{\mrightb + \mleftb}{1 + \mrightb \mleftb}
\end{equation}
and variance 
\begin{equation} \label{eqn:sigmatrue}
\sigmatrue = 1 - (\mtrue)^2,
\end{equation}
where $\mleftb$ is the mean of $\muleftb$ (\ref{eqn:muleftb}), 
which is formed as in (\ref{eqn:MeanOfMrightb}). 
The approximate global function (corresponding to (\ref{eqn:ApproxGraph})) is 
the Gaussian
\begin{equation} \label{eqn:murightg_muleftg}
\murightgM(x) \muleftg(x)
\end{equation}
with mean $\mg$ and variance $\sigmag$ given by
\begin{align}
1/\sigmag      &\eqm{=}  1/\sigmarightgM + 1/\sigmaleftg  \label{eqn:sigmag}\\
\mg / \sigmag  &\eqm{=}  \mrightgM / \sigmarightgM + \mleftg / \sigmaleftg.  \label{eqn:mg}
\end{align}

Now a natural choice for the approximation (\ref{eqn:GenRuleApprox}) is to 
equate the mean and the variance of the Gaussian approximation with the corresponding moments 
of the true global function: 
\begin{equation} \label{eqn:MinkaGaussianApprox}
\mg = \mtrue \text{~~~and~~~} \sigmag = \sigmatrue.
\end{equation}
(As pointed out by Minka, 
this choice may be derived from~(\ref{eqn:MinDivergence}).)
The desired Gaussian message $\murightgM$ is thus obtained by 
first evaluating (\ref{eqn:mtrue}) and (\ref{eqn:sigmatrue}) and then computing 
$\sigmarightgM$ and $\mrightgM$ from (\ref{eqn:sigmag}) and (\ref{eqn:mg}).

Note that, in general, the message $\murightgM$ is not trivial 
even if $\murightb$ is neutral ($\mrightb=0$ and $\sigmarightb=1$).

\section{Issues}
\label{sec:Issues}

\subsection{Negative ``Variance''}

Solving (\ref{eqn:sigmag}) for $\sigmarightgM$ may result in a negative value for $\sigmarightgM$. 
(This indeed happens in the examples to be described in Section~\ref{sec:SimulationResults}.) 
In such cases, $\murightgM$ is a correction factor 
(not itself a probability mass function) 
that tries to compensate for an overly confident $\muleftg$. 
The product (\ref{eqn:murightg_muleftg}) usually remains a valid probability mass function, 
up to a scale factor.

\subsection{Damping}

In our numerical experiments (Section~\ref{sec:SimulationResults}), 
simply replacing the standard Gaussian message $\murightgs$ 
by $\murightgM$ yielded unstable algorithms. 
Good results were obtained, however, by geometric mixtures of the form 
\begin{equation} \label{eqn:murightmixed}
\murightg(x) = \left(\murightgM(x)\right)^{\alpha} \left(\murightgs(x)\right)^{1-\alpha}
\end{equation}
with $0\leq \alpha \leq 1$. 
The mean and the variance of the resulting Gaussian $\murightg$ are given by
\begin{equation} \label{eqn:mixedsigma}
1 / \sigmarightg = \alpha / \sigmarightgM + (1-\alpha) / \sigmarightgs
\end{equation}
and
\begin{equation} \label{eqn:mixedmean}
\mrightg = \frac{\mrightgM \alpha/\sigmarightgM + \mrightgs (1-\alpha)/\sigmarightgs}%
                {\alpha / \sigmarightgM + (1-\alpha) / \sigmarightgs}
\end{equation}
\vspace{0.5ex}

\section{Application Example: Equalization}
\label{sec:SimulationResults}
Consider the transmission of binary ($\{+1,-1\}$-valued) symbols 
$X_1,\ldots,X_n$ over a linear channel with transfer function 
$H(z) = \sum_{\ell=0}^M h_\ell z^{-1}$ and additive white Gaussian noise $W_1,\ldots,W_n$. 
The received channel output symbols are $Y_1,\ldots,Y_n$ with
\begin{equation} \label{eqn:TransmissionModel}
Y_k = \sum_{\ell=0}^M h_\ell X_{k-\ell} + W_k,
\end{equation}
where we assume $X_k=0$ for $k<0$. The binary symbols $X_k$ may or may not be coded.

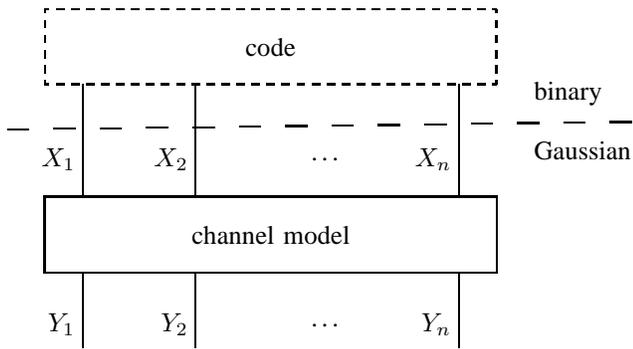
\begin{figure}
\begin{center}
\begin{picture}(75,45)(0,0)
\put(0,35){\dashbox(60,10){code}}
\dashline{3}(-5,29)(78,30)    \put(65,33){binary}
                              \put(65,25){Gaussian}
\put(5,20){\line(0,1){15}}    \put(4,25){\pos{r}{$X_1$}}
\put(20,20){\line(0,1){15}}   \put(19,25){\pos{r}{$X_2$}}
\put(37.5,25){\cent{\ldots}}
\put(55,20){\line(0,1){15}}   \put(54,25){\pos{r}{$X_n$}}
\put(0,10){\framebox(60,10){channel model}}
\put(5,0){\line(0,1){10}}     \put(4,3){\pos{r}{$Y_1$}}
\put(20,0){\line(0,1){10}}    \put(19,3){\pos{r}{$Y_2$}}
\put(37.5,3){\cent{\ldots}}
\put(55,0){\line(0,1){10}}    \put(54,3){\pos{r}{$Y_n$}}
\end{picture}
\end{center}
\caption{\label{fig:CodeChannelGraph}%
Joint code/channel factor graph.}
\end{figure}

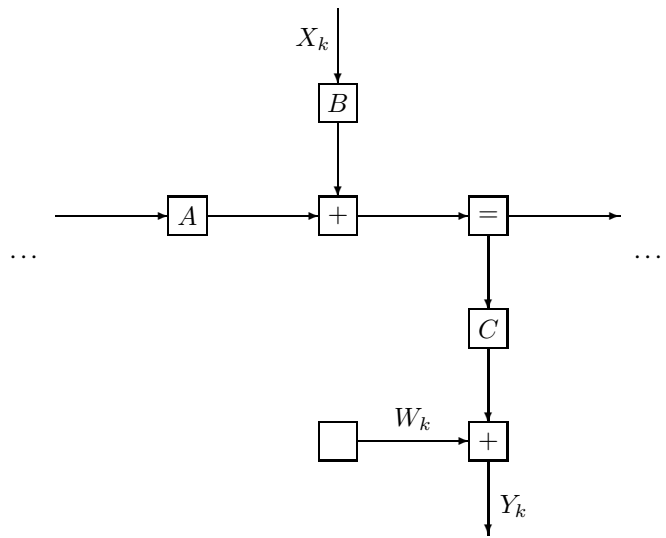
\begin{figure}
\begin{center}
\begin{picture}(75,70)(0,0)
%
\put(-4,37){\cent{\ldots}}
\put(0,42.5){\vector(1,0){15}}
\put(15,40){\framebox(5,5){$A$}}
\put(20,42.5){\vector(1,0){15}}
\put(35,40){\framebox(5,5){$+$}}
\put(37.5,70){\vector(0,-1){10}}      \put(36.5,66){\pos{r}{$X_k$}}
\put(35,55){\framebox(5,5){$B$}}
\put(37.5,55){\vector(0,-1){10}}
\put(40,42.5){\vector(1,0){15}}
\put(55,40){\framebox(5,5){$=$}}
\put(60,42.5){\vector(1,0){15}}
\put(79,37){\cent{\ldots}}
\put(57.5,40){\vector(0,-1){10}}
\put(55,25){\framebox(5,5){$C$}}
\put(57.5,25){\vector(0,-1){10}}
\put(55,10){\framebox(5,5){$+$}}
\put(57.5,10){\vector(0,-1){10}}      \put(59,3){$Y_k$}
\put(35,10){\framebox(5,5){}}
\put(40,12.5){\vector(1,0){15}}       \put(47.5,15.5){\cent{$W_k$}}
\end{picture}
\end{center}
\caption{\label{fig:ChannelStateSpaceGraph}%
Factor graph of the channel model (one section).}
\end{figure}

The joint code/channel factor graph is shown in \Fig{fig:CodeChannelGraph} 
with channel-model details as in \Fig{fig:ChannelStateSpaceGraph}. 
(In the uncoded case, the code graph is missing.) 
The factor graph shown in \Fig{fig:ChannelStateSpaceGraph} results 
from writing (\ref{eqn:TransmissionModel}) in state space form with 
suitable matrices $A$, $B$, and $C$, 
where $B$ is a column vector and $C$ is a row vector, cf.\ \cite{Lg:ifg2004}. 

Equalization is achieved by forward-backward Gaussian message passing 
(i.e., Kalman smoothing)
in the factor graph of \Fig{fig:ChannelStateSpaceGraph} according to the 
recipes stated in \cite{Lg:ifg2004}. 
(See \cite{Lg:Turbo2006c} for a more detailed discussion.) 

In this paper, we are only concerned with the messages along the edges $X_k$ 
(towards the channel model) in \Fig{fig:CodeChannelGraph}.
Using the standard messages (\ref{eqn:StandardGaussianApprox}) 
results in an LMMSE equalizer; in the uncoded case, 
this algorithm terminates after a single forward-backward sweep 
since the factor graph of \Fig{fig:ChannelStateSpaceGraph} has no cycles. 
However, using the (damped) Minka messages (\ref{eqn:murightmixed})--(\ref{eqn:mixedmean}) 
results in an iterative algorithm even in the uncoded case. 

\begin{figure}
\begin{center}
\includegraphics[width=85mm]{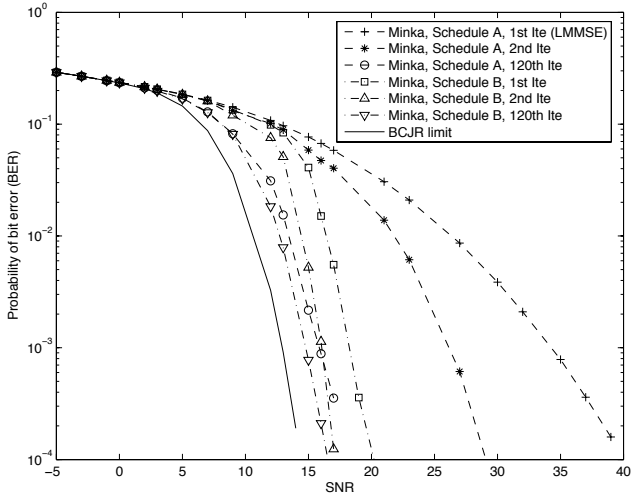}
\end{center}
\vspace{-2mm}
\caption{\label{fig:FIR5_uncoded}%
Bit error rate vs.\ SNR
for uncoded binary transmission over FIR channel with transfer function 
$H(z) = 0.227 + 0.46 z^{-1} + 0.688 z^{-2} + 0.46 z^{-3} + 0.227 z^{-4}$.}
\end{figure}

\begin{figure}
\begin{center}
\includegraphics[width=85mm]{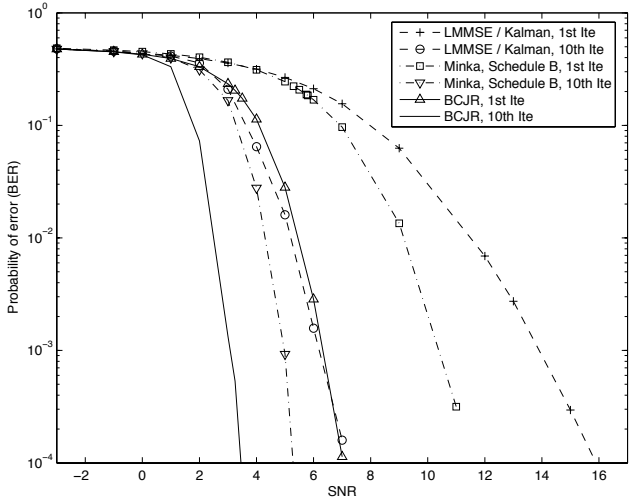}
\end{center}
\vspace{-2mm}
\caption{\label{fig:FIR5_coded}%
Bit error rate vs.\ SNR
for coded binary transmission over FIR channel.}
\end{figure}

\begin{figure}
\begin{center}
\includegraphics[width=85mm]{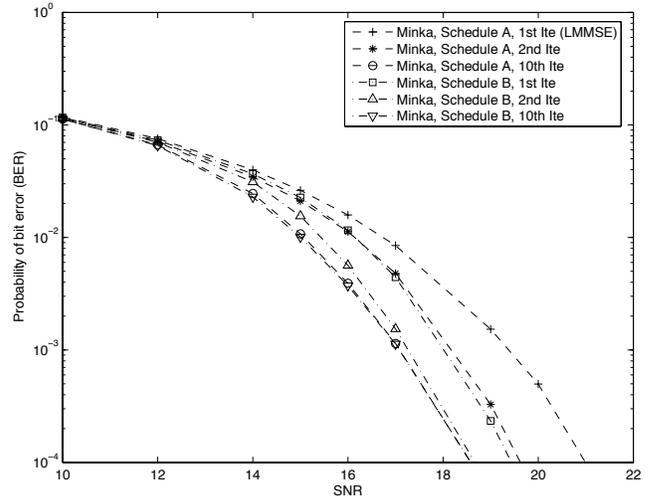}
\end{center}
\vspace{-2mm}
\caption{\label{fig:IIR1_uncoded}%
Bit error rate vs.\ SNR
for uncoded binary transmission over IIR channel with transfer function
$H(z) = 1/(1 - 0.9 z^{-1})$.}
\end{figure}

Simulation results for two different channels are given in 
Figures \ref{fig:FIR5_uncoded}--\ref{fig:IIR1_uncoded}. 
Figures \ref{fig:FIR5_uncoded} and~\ref{fig:FIR5_coded} 
show the bit error rate vs.\ the signal-to-noise ratio (SNR) 
for an FIR channel with transfer function 
$H(z) = 0.227 + 0.46 z^{-1} + 0.688 z^{-2} + 0.46 z^{-3} + 0.227 z^{-4}$;
\Fig{fig:IIR1_uncoded} shows the bit error rate vs.\ the SNR 
for an IIR channel with transfer function $H(z) = 1/(1 - 0.9 z^{-1})$. 
The FIR channel was used as an example in \cite{Pro:dc};
because this channel has a spectral null, the difference between 
a LMMSE equalizer and the optimal BCJR equalizer is large. 
The IIR channel was used as an example in \cite{DHH:ddfse1989}. 

Two different message update schedules are used:
in Schedule~A, the output messages (along edge $X_k$ out of the channel model) 
are initialized to ``infinite'' variance and are updated only after 
a complete forward-backward Kalman sweep;
in Schedule~B, these messages are updated  
(and immediately used for the corresponding incoming Minka message) 
both during the forward Kalman sweep and the backward Kalman sweep. 
From our simulations, Schedule~B is clearly superior. 

It is obvious from Figures \ref{fig:FIR5_uncoded} and~\ref{fig:IIR1_uncoded} 
that, for uncoded transmission, the Minka messages provide 
a very marked improvement over the standard messages (i.e., over the LMMSE equalizer). 
In \Fig{fig:FIR5_uncoded}, we almost achieve the performance of the BCJR (or Viterbi) 
equalizer (and we also outperform the decision-feedback equalizer \cite[p.~643]{Pro:dc}).
As for \Fig{fig:IIR1_uncoded}, we almost achieve the performance of 
the quasi-Viterbi algorithm reported in \cite{DHH:ddfse1989}. 

For the coded example of \Fig{fig:FIR5_coded}, a rate 1/2 convolutional codes 
with constraint length~7 was used. In this case, the iterative Kalman equalizer 
does quite well already with the standard input messages (\ref{eqn:StandardGaussianApprox}),  
but the Minka messages do give a further improvement at very small cost. 

\begin{figure}
\visual\vspace{5mm}
\begin{center}
\includegraphics[width=80mm]{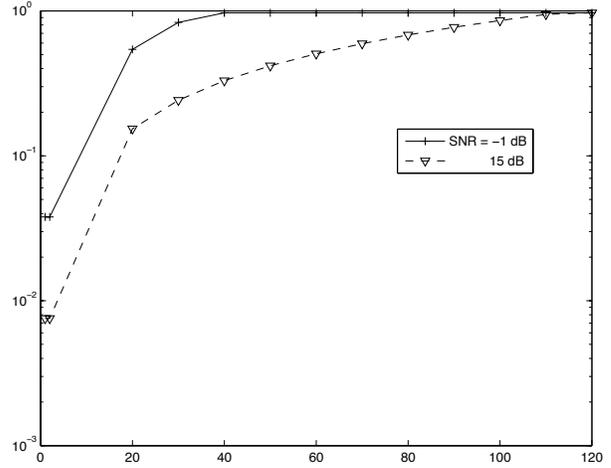}
\end{center}
\caption{\label{fig:alpha}%
Good sequences for $\alpha$ vs.\ the iteration number $k$.}
\end{figure}

A key issue with all these simulations is the choice of the damping/mixing factor $\alpha$ 
in (\ref{eqn:murightmixed})--(\ref{eqn:mixedmean}). The best results were obtained 
by changing $\alpha$ in every iteration. 
Typical good sequences of values of $\alpha$ are plotted in \Fig{fig:alpha}. 
We note the following observations:
\begin{itemize}
\item
The initial values of $\alpha$ are very small. 
\item
After a moderate number of iterations (typically 10\ldots 20), 
the bit error rate 
stops decreasing. 
At this point, $\alpha$ is still very small. 
\item
Many more iterations with slowly increasing $\alpha$ are required to 
reach a fixed point with $\alpha=1$.
\item
At such a fixed point with $\alpha=1$, the approximation~(\ref{eqn:GenRuleApprox}) 
holds everywhere. 
\end{itemize}

\section{Conclusion}
Elaborating on Minka's work, 
we have formulated a general computation rule for lossy messages. 
An important special case is the conversion of ``soft-bit'' messages to Gaussian messages. 
In this case, the resulting Gaussian message is non-trivial even if the ``soft-bit'' message 
is neutral. 
By this method, the performance of a Kalman equalizer is significantly improved.


\newcommand{\COM}{IEEE Trans.\ Communications}
\newcommand{\IT}{IEEE Trans.\ Information Theory}
\newcommand{\JSAC}{IEEE J.\ on Selected Areas in Communications}
\newcommand{\SPMag}{IEEE Signal Proc.\ Mag.}

\end{document}